\documentclass[twocolumn,pre,floats,aps,amsmath,amssymb,nofootinbib]{revtex4-2}
\usepackage{graphicx}
\usepackage{bm}
\usepackage{physics}
\usepackage[labelformat=empty]{caption}

\begin{document}

\title{The Pseudoscience of Time Travel}
\author{Andrew Knight, J.D.}
\affiliation{aknight@alum.mit.edu}
\date{\today}

\begin{abstract}
Because closed timelike curves are consistent with general relativity, many have asserted that time travel into the past is physically possible if not technologically infeasible.  However, the possibility of time travel into the past rests on the unstated and false assumption that zero change to the past implies zero change to the present.  I show that this assumption is logically inconsistent; as such, it renders time travel into the past both unscientific and pseudoscientific.
\end{abstract}

\maketitle

\section{Introduction}
\label{sec:intro}

The possibility of time travel into the past is a staple of science fiction but is also taken seriously by the physics academy.  For instance, “closed timelike curves,” the physicist’s phrase for time travel, are discussed widely in the physics literature \cite{Pienaar,Godel,Morris,Deser,Deutsch,Hartle,Politzer,Goldwirth}.

Far from relegating time travel to science fiction, many celebrated physicists discuss it seriously both as being subject to scientific inquiry and as being physically possible, perhaps even technologically feasible in the not-too-distant future.  For example, Stephen Hawking \cite{Hawking2} asserted that quantum theory “[should] allow time travel on a macroscopic scale, which people could use.”\footnote{A skeptic of the possibility of time travel, he articulated an unproven “chronology protection conjecture” \cite{Hawking1} that would prevent it.}  Nobel Prize winner Kip Thorne \cite{Thorne} asserts that “If wormholes can be held open by exotic material, then [possibilities for time travel into the past] are general relativity’s predictions.”  Crucially, he conjectures that time travel is simply a matter of technological feasibility as there are “no unresolvable paradoxes” in time travel.

I will argue in this paper that the possibility of time travel into the past rests on a logical contradiction.  As such, it is not merely false but, more importantly, pseudoscientific as improperly treated as subject to scientific inquiry.

\section{Pseudoscience}

What is and is not scientific has been subject to debate since time immemorial.  For example, Karl Popper \cite{Popper} argues that a statement or hypothesis is scientific if and only if it is falsifiable, while others argue that anything empirically testable is scientific.  Moreover, what constitutes \textit{pseudoscience} is subject to further debate.\footnote{See, for example, https://plato.stanford.edu/entries/pseudo-science/.}  A false statement need not be pseudoscientific.  For example: “Gravity causes apples to fall upward” is a statement that is in contradiction with empirical evidence; as such, it is both testable and falsifiable.  Indeed, most scientific hypotheses over the past millennia have been falsified.  Such hypotheses might have been true but turned out to conflict with the facts of the universe.

However, a statement that is internally inconsistent – i.e., self-contradictory – is inherently false.  A statement that is logically contradictory has no chance of being confirmed or falsified through experimentation and is therefore not subject to scientific inquiry.  Rather, a contradiction is false by logical necessity.  The statements “1=2,” “The sky is blue and it is not blue,” and $A \cap \neg A$ are necessarily false.  As contradictions, they are not testable, falsifiable, or scientific.  A logical contradiction arising in an inquiry renders that inquiry unscientific, and -- to the extent that inquiry masquerades as science -- pseudoscientific.

The soundness of logical arguments depends both on the veracity of their assumptions and the validity of their logical algorithms.  Often, however, an ostensibly sound logical argument may depend on an unidentified assumption that is either false or in contradiction with the argument’s conclusion.

For example, if the truth of premises $A$ and $B$ is sufficient to guarantee the truth of statement $D$, then $(A \cap B \Rightarrow D)$ is a true statement.  However, if the truth of $D$ depends on the truth of another premise $C$, then it must be the case that $(D \Rightarrow C)$.  This is because the situation in which $(A \cap B \Rightarrow D)$ is false but $(A \cap B \cap C \Rightarrow D)$ is true happens when $(\neg C \Rightarrow \neg D)$, which is the contrapositive and equivalent of $(D \Rightarrow C)$.

In the case of time travel $TT$ into the past, the most common assertion of its possibility looks, in heavily simplified form, something like $(GR \cap W \Rightarrow TT)$, where $GR$ refers to the truth of Einstein's general relativity and $W$ refers to the possibility of producing an adequately large and stable wormhole connecting two distant points in spacetime.  However, independently of whether premises $GR$ and $W$ are true, the implication may be false if it depends on some unindentified assumption or condition $C$.  If so, then $(TT \Rightarrow C)$ and, equivalently, ($\neg C \Rightarrow \neg TT$).

If the current narrative among the physics community regarding the possibility of time travel indeed depends on some unidentified assumption that can be shown false on purely \textit{a priori} logical grounds, then that narrative is not merely false; it is also pseudoscientific.

\section{Unidentified Assumption}

In nearly every account of time travel in which some aspect of the past is changed, it is implicitly assumed that a corresponding or commensurate change would occur to the present.  Certainly, chaotic amplification – i.e., the so-called ``butterfly effect" – would likely yield larger and more unexpected changes to the present than we might hope.\footnote{Indeed, changing the initial conditions of three gargantuan black holes by an amount smaller than the Planck length would still, given enough time, render their evolutions entirely unpredictable \cite{Boekholt}.}  Still, it is tempting to believe that changes to the present depend only on changes to the past, and the \textit{magnitude} of changes to the present would be commensurate with the \textit{magnitude} of changes to the past.  

For example, the so-called grandfather (\textit{aka} matricide) paradox, in which a time traveler kills one of his ancestors, is widely regarded as a physical limitation to the possibility of time travel.  Further, because of the sensitivity of nonlinear dynamics to the precision of initial conditions, it is often (though not always) recognized that even seemingly insignificant changes to the past could result in temporal paradoxes.  Hawking \cite{Hawking2}, among others, asserted that such a paradox can be resolved, and the possibility of time travel into the past maintained, as long as “when you did go back [in time], you wouldn’t be able to change recorded history.”

This popular but fundamentally errant solution relies on an important unstated assumption.  Let $\Delta (t_0<t_1)$ be the change to the state of the universe at time $t_0$ caused by a time traveler traveling back from time $t_1$, and let $\Delta (t_1)$ be the change to the state of the universe at time $t_1$ caused by change $\Delta (t_0<t_1)$.  The unstated (and, as I will argue, fatal) assumption underlying any scientific treatment of time travel is therefore:
\begin{equation}
\lim_{\Delta (t_0<t_1) \to 0} \Delta (t_1) = 0
\end{equation}

Further, it is assumed that $\Delta (t_0<t_1)$ can be made arbitrarily small with the result that $(\Delta (t_0<t_1) = 0) \Rightarrow (\Delta (t_1) = 0)$.  It must be stressed that if time travel to the past is possible, then Eq.\ 1 must be true.  That is, if time travel is to be possible at all, then its likelihood and technological ease must increase with fewer changes made to the past, with the best prospects for time travel achieved with zero changes to the past.

Indeed, if Eq.\ 1 is correct, then a hopeful time traveler might simply plan to travel back in time and make as few changes as possible – ideally zero changes – to prevent the possibility of changes to the present (including any accompanying time paradox).  After all, even the tiniest measurable change to the past might chaotically amplify -- \textit{a la} the grandfather paradox -- in a way that prevents the time traveler from choosing to time travel, or even from existing.  In other words, if the hopeful time traveler could be certain to make exactly no changes to the past, then he would presumably be guaranteed no changes to the present.

Is this notion tenable?  No.  First, being present in any form in the past requires making physical changes that would chaotically amplify over time.  Even passive observation requires the absorption of photons by one’s retinas.  It is hard to imagine how one might time travel into the past and perceive any aspect of that experience without interacting in some physical form with his surroundings, and these interactions would necessarily influence the future.  Second, as I will argue below, Eq.\ 1 is false.

\section{False Assumption}

Assume Alice’s watch currently says time $t_2$.  She has a record of a quantum interference experiment in which at earlier time $t_0$, an object $O$ is in a quantum superposition state $(\ket{O}=x\ket{X}+y\ket{Y})$, with $\ket{X}$ corresponding to semiclassical localization at position X and $\ket{Y}$ corresponding to semiclassical localization at position Y, where $x$ and $y$ are complex amplitudes.  She also has a record that at time $t_1$ (where $t_0<t_1<t_2$), the object was actually detected at position X.  

Assume it is possible for Alice to travel backward to time $t_0$ to later witness the outcome of the quantum interference experiment.  She intends to make no changes at all, particularly to the experiment.\footnote{This is another untenable assumption.  Just as her very presence at time $t_0$ influences her environment via her body's interactions with photons and air molecules, the timelike separation between her presence at $t_0$ and the detection event at $t_1$ requires, quantum mechanically, that she \textit{will} affect the experiment, for two reasons.  First, in a process limited by the speed of light, there is some finite probability that one or more particles that Alice's body has affected will then affect the object's superposition state prior to detection at position X or Y.  Second, in a process not limited by the speed of light, particles affected by Alice's body may already be entangled with the object and/or its measuring device such that changes to those particles instantaneously change the quantum amplitudes relevant to the experiment.}  What is the state of the object $O$ when she “arrives” at time $t_0$?  There are only two possibilities:
\begin{itemize}
\item A)	She will with certainty detect the object at position X at time $t_1$.
\item B)	She will \textit{not} with certainty detect the object at position X at time $t_1$.
\end{itemize}

If A) is true, then the object must have been in eigenstate $\ket{X}$ at $t_0$, not in superposition state $(x\ket{X}+y\ket{Y})$.  That implies that she \textit{did} change the experiment (as well as the past in general).  Therefore A) is false.

If B) is true, then it is possible that she may detect the object at position Y at time $t_1$.  If so, there are only two possibilities:
\begin{itemize}
\item C)	She detects the object at position Y at time $t_1$.
\item D)	She detects the object at position X at time $t_1$.
\end{itemize}

If C) is true, then her time travel changes the past, despite her best efforts, even if she has not physically interacted in any way with past events.  If D) is true, then it is merely an accident having probability $|x|^2$.  

In other words, even if Alice makes no changes to the past, her present will indeed change with a probability $p \geq 1 - |x|^2$.  But how many quantum events happen between times $t_1$ and $t_2$?  For Alice to succeed in making no changes to the present, every one of those events must resolve in exactly the same way as they are recorded in the information structure of her present universe.

For instance, consider that a second object, independent of the first object and in state $(\ket{O_2}=w\ket{W}+z\ket{Z})$ at time $t_0$, had been detected at position Z at time $t_1$.  If Alice's time travel back to $t_0$ is to have no effect on her present, then her later detection of the second object at position Z will be a mere accident having probability $|z|^2$.  Because both experiments must resolve in the same way, their joint probability is simply their product $|x|^2 |z|^2$.  In general, if N quantum events happen within the shared light cones of spacetime events at times $t_1$ and $t_2$, the probability of these events resolving in a particular manner decreases exponentially with N.  For all practical purposes, the probability is zero.  

Therefore, time travel into the past, even with ostensibly zero changes to the past, would guarantee changes to the present.  Eq.\ 1 is false.  Because the possibility of time travel into the past depends on the unidentified assumption that Eq.\ 1 is true, time travel is not possible.  More importantly, Eq.\ 1 has been shown false on purely logical grounds\footnote{One might object that the argument depends on the truth of quantum mechanics, although all serious proposals for time travel and closed timelike curves in the physics literature take the veracity of quantum mechanics as a given anyway.}, which means that the possibility of time travel is inherently self-contradictory and is not subject to empirical verification or falsification.  To the extent that it continues to be treated as the subject of scientific inquiry, time travel is pseudoscience.

\section{Objection}

One might object that the argument in this paper rests on the assumption that quantum mechanics implies indeterminism.  Proponents of a deterministic block universe, including those who hold the Many Worlds Interpretation (``MWI") of quantum mechanics and those who assert that the quantum wave state of every system always evolves linearly and reversibly, might argue that both statements C) and D) are true, with $Alice_X$ detecting the object at position X and $Alice_Y$ detecting it at position Y.  Eq.\ 1 is essentially an assertion of determinism, that if no changes are made to the past then no changes will be made to the present.  According to MWI, Alice may indeed find herself in a different \textit{branch} of infinitely many present universes, but the present multiverse itself will not change if she makes no changes to the past.  

This objection fails.  For instance, what would it mean for Alice to find herself in a branch of the multiverse in which, due to chaotic amplification of some tiny interaction, she decides against time travel, or isn't born?  Rather, the point of Section IV is that Alice cannot travel into the past without changing what \textit{she perceives} as her present universe.  Because proposals for time travel all inherently depend on the applicability of Eq.\ 1 to the time traveler's own perspective, these proposals fail and fall outside the scope of scientific inquiry. 

\section{Conclusion}

When we assume that zero change to the past implies zero change to the present, we impose the information structure of the present universe onto its past structure – that is, we assume that everything will turn out the same, except for those events (and their chaotic interactions) that were changed in the past.  Unfortunately for aspiring time travelers, this notion is false.  

Traveling into the past logically requires changing the present, no matter how careful one is to avoid a temporal paradox.  Hopes to travel into the past without changing the present, such as by avoiding any physical interaction within the past, are unfounded.  Because all serious proposals for time travel into the past inherently assume Eq.\ 1, which has been shown false on \textit{a priori} logical grounds, their treatment as scientific proposals are pseudoscientific.

\end{document}